# Role of surface functional groups to superconductivity in Nb$_2$C-MXene: Experiments and density functional theory calculations

Kai Wang[1#], Haolin Jin[1#], Hongye Li[1], Zhongquan Mao[1], Lingyun Tang[1], Dan Huang[2], Ji-Hai Liao[1*] & Jiang Zhang[1*]

[1]*Department of Physics, South China University of Technology, Guangzhou, 510640, China*

[2]*MOE Key Laboratory of New Processing Technology for Non-ferrous Metals and Materials, Guangxi Key Laboratory of Processing for Non-ferrous Metals and Featured Materials, Guangxi University, Nanning, 530004, China*

[#]These authors contributed equally: Kai Wang and Haolin Jin

[*]Corresponding author. Emails: jhliao@scut.edu.cn (J-H. L.) and jonney@scut.edu.cn (J. Z.)

**Abstract**: The recently discovered surface-group-dependent superconductivity in Nb$_2$C-MXene fabricated by the molten salts method is attracting wide attention. However, regarding the superconductivity of Nb$_2$C-MXene with functional F groups (Nb$_2$CF$_x$), there were some conflicting results in experimental and theoretical studies. Herein, we systematically carried out experimental and theoretical investigations on the superconductivity in Nb$_2$C-MXene with the Cl functional group (Nb$_2$CCl$_x$) and Nb$_2$CF$_x$. The experimental results of the Meissner effect and zero resistivity have proved that Nb$_2$CCl$_x$ is superconducting with the transition temperature ($T_c$) ~ 5.2 K. We extract its superconducting parameters from the temperature dependence of resistivity and the field dependence of the magnetization. The Ginzburg-Landau parameter $\kappa_{GL}$ is estimated to be 2.41, indicating that Nb$_2$CCl$_x$ is a typical type-II superconductor. Conversely, both magnetic and electrical transport measurements demonstrate that Nb$_2$CF$_x$ is not superconducting. The first-principles density functional theory (DFT) calculations show that the $T_c$ of Nb$_2$CCl$_x$ is ~ 5.2 K, while Nb$_2$CF$_x$ is dynamically unstable with imaginary frequency in phonon spectrum, which is in good agreement with the experimental results. Our studies not only are useful for clarifying the present inconsistency but also offer referential significance for future investigations on the superconductivity of MXenes.

# 1. Introduction

MXenes are generally referred as a new type of two-dimensional (2D) transition metal carbide, nitride and boride with chemical formula of $M_{n+1}X_nT_x$, where M represents transition metal, X represents carbon, nitrogen, or boron, n=1, 2, 3, T is surface-linked F, OH, O, Cl, and other active functional groups [1-6]. With unique properties such as good electrical conductivity, magnetic properties, and thermoelectric properties, MXenes are quite useful in a large number of applications, including energy storage, optoelectronic, biomedical, electromagnetic shielding and photovoltaic [7-14].

In terms of superconductivity, $α$-$Mo_2C$ crystal with the transition temperature ($T_c$) of 3.6 K was the first reported two-dimensional (2D) transition metal carbide, which showed observable characteristics of Berezinskii-Kosterlitz-Thouless transition [15,16]. But $α$-$Mo_2C$ is different from the MXenes we mentioned above, it was grown using the chemical vapor deposition (CVD) technique, not etched from the MAX phase, so it has no functional group [17]. Although fewer superconductors of the MAX phase, such as $Nb_2AlC$ ($T_c$ ~ 0.44 K), $Lu_2SnC$ ($T_c$ ~ 5.2 K), $Nb_2InC$ ($T_c$ ~ 7.5 K) [18-20], were reported previously. It was only recently discovered that the $Nb_2C$-MXene with different functional groups ($Nb_2CCl_2$ ($T_c$ ~ 6 K), $Nb_2CS_2$ ($T_c$ ~ 6.4 K), $Nb_2CSe$ ($T_c$ ~ 4.5 K), and $Nb_2C(NH)$ ($T_c$ ~ 7.1 K)) were superconducting, which were fabricated by performing substitution and elimination reactions in molten inorganic salts [21]. Besides, the superconducting parameter, such as $H_{c2}$, showed a strong dependence on the surface functional group, the $H_{c2}$ of $Nb_2CNH$ was nearly 3 times that of $Nb_2CCl_2$ [21], which suggested that surface groups played a crucial role to the superconductivity of $Nb_2C$ MXene. Through the magnetic measurement, $Nb_2C$-MXene with the F functional group ($Nb_2CF_x$) etched by hydrofluoric acid (HF) was found to show the Meissner effect with the highest $T_c$ of 12.5 K [22, 23]. However, zero resistivity, which is identified as the most important evidence of superconductivity, was absent. The temperature dependence of resistivity for the pressed $Nb_2CF_x$ pellet indicated a sharp increase with decreasing temperature [21], which is similar to the electrical properties of $Nb_2CT_z$-yLi thin film [24]. Neither two transport experiments of $Nb_2CF_x$ MXene did show the superconducting behavior. On the other hand, first-principle calculations on intrinsic $Nb_2C$ showed no superconductivity [25]. Theoretical studies on both pristine and functionalized $Nb_2C$-MXene using the density functional theory (DFT) suspected that the possible reason for the superconductor with $T_c \sim 12.5$ K observed is $Nb_2CO_2$ rather than $Nb_2CF_x$ [26].

To clarify the above confused and inconsistent results and in-depth understand the superconductivity of $Nb_2C$-MXene, we have systematically studied the superconductivity of $Nb_2CCl_x$ and $Nb_2CF_x$ through experiments and theoretical calculations. We prepared $Nb_2CCl_x$ and $Nb_2CF_x$ respectively and studied the magnetic and electrical transport properties as a function of temperature. Combined with the theoretical computations, our results show that different functional groups do have a great influence on the superconductivity of $Nb_2C$-MXene, when the linked functional group is Cl the superconducting performance is better, and when the linked functional group is F, there is no superconductivity was observed.

**2. Experimental and Characterization**

2.1 Preparation of $Nb_2CF_x$ and $Nb_2CCl_x$

The materials used here include hydrochloric acid solution (wt 38%), lithium fluoride powder (LiF, purity 97%), hydrofluoric acid solution (wt 40%), niobium aluminum carbon powder ($Nb_2AlC$, purity 97%, 400 mesh), anhydrous cadmium chloride powder ($CdCl_2$, purity 99%), deionized water and anhydrous ethanol.

For $Nb_2CF_x$, we employed two methods to strip the $Nb_2AlC$ MAX phase. Hydrothermal method [27]: Firstly, 2 g lithium fluoride powder and 40 ml hydrochloric acid solution were mixed and sonicated for 30 mins. Then 0.5 g $Nb_2AlC$ powder was added to the mixed solution. The solution was sealed in an autoclave and heated at 120 °C for 36 h. HF etching method: 0.5 g $Nb_2AlC$ powder was added to 40 ml of hydrofluoric acid solution and reacted for 96 h in a tetrafluoroethylene reactor at 60 °C.

For $Nb_2CCl_x$, we used the molten salt method to etch $Nb_2AlC$ [21]. Firstly, the $CdCl_2$ powder and the $Nb_2AlC$ powder were mixed according to a molar ratio of 10:1 and ball milled for 12 hours. The mixture was put in a corundum crucible and kept at 300 °C for 8 hours in a tube furnace under the flow of argon gas (mixed 5% hydrogen), and then reacted at 750 °C for 36 hours. After the completion of the reaction, the obtained products were stirred and washed with hydrochloric acid for 12 hours.

We adopted the same post-processing process to get the MXene products. The reacted solution was centrifuged at 7000 rpm for 5 min and washed with deionized water several times ultrasonically until the pH of the supernatant reached 7. The precipitate was taken out from the centrifuge tube with ethanol and dried at 55 ℃ under vacuum for 12 hours to obtain $Nb_2C$-MXene powders.

2.2 Characterization

The structure and morphology of $Nb_2C$-MXene were characterized using scanning electron microscopy (SEM, Gemni500) and X-ray diffraction (XRD, Bruker D8 ADVANCE). DC and AC magnetic susceptibility, as well as resistivity, were measured by the Physical Property Measurement System (PPMS Evercoll II, Quantum Design) in the temperature range of 2-300 K with a vibrating sample magnetometer and electrical transport option. The powders samples were pressed into a circular disc with a diameter of 8 mm and a thickness of 1.5 mm, and the standard four-wire method was used to measure resistivity. The chemical composition of $Nb_2C$-MXene was defined by an energy dispersive X-ray (EDX, OXFORD X-Max[N]) and X-ray photoelectron spectrometer (XPS, Thermo Fisher Scientific K-Alpha).

2.3 Calculation details

Our calculations are based on DFT used in the QUANTUM-ESPRESSO package [28]. We use ultra-soft pseudopotential with generalized gradient approximation (GGA) in Perdew-Burke-Ernzerhof (PBE) format for exchange correlation potential [29, 30]. The *4s*, *5s*, *4p*, and *4d* electrons of Nb, the *2s* and *2p* electrons of C and F, and the *3s* and 3*p* electrons of Cl are considered as valence electrons.The threshold energy of a plane wave is defined as 680 eV (50 Ry), and all crystal structures are completely relaxed until the Herman-Feynman force on each atom is less than $2.57\times10^{-4}$ eV/Å ($10^{-5}$ Ry/Bohr). A Methfessel-Paxton [31] smearing of 0.272 eV (0.02 Ry) was used for the corresponding electronic self-consistent cycles. Phonon frequencies and EPC parameter $\lambda$ were calculated with the phonon wave-vector meshes of 6×6×1 and the denser k meshes of 24×24×4.

## 3. Results and discussion

### 3.1 Structure and morphology

The XRD patterns of $Nb_2CF_x$ MXene obtained from the hydrothermal method and $Nb_2CCl_x$ are illustrated in Fig. 1a and b. These two profiles can be indexed to *P6₃/mmc* with lattice parameters of $a$ = 3.134 Å, $c$ = 23.68 Å and $a$ = 3.162 Å, $c$ = 17.6551 Å, respectively. From the comparison of the XRD pattern of the $Nb_2AlC$ MAX phase (Fig. S1a), we can find that the (102) peak of the MAX phase near 39° have disappeared. At the same time, the shift of the (002) peak to a low angle proves that the layer spacing has increased significantly after etching. The position of the (002) diffraction peak of $Nb_2CF_x$ has changed from 12.8° to 7.37° and the *d*-spacing has increased from 6.91 Å to 11.98 Å. Meanwhile, the position of the (002) diffraction peak of $Nb_2CCl_x$ has changed to 10.03° and the *d*-spacing has increased to 8.82 Å. The shift of the (002) peak and the disappearance of the remaining peaks confirm the conversion of the MAX phase to the MXene phase after etching[22]. The SEM images of $Nb_2CF_x$ and $Nb_2CCl_x$ are illustrated in Fig. 1c and d. Both of the MXene products have a typical layered structure. XRD pattern and SEM image of $Nb_2CF_x$ obtained from HF etching present a similar structure (Fig S1b, c).

We used XPS and EDX to characterize the contents of F and Cl in $Nb_2CF_x$ and $Nb_2CCl_x$ as shown in Fig. S2. The existence of F and Cl is evidenced by the main peak at 685.6 eV and 199.5 eV, respectively. It can be seen that Al is basically etched away from the results of the elemental composition measured by EDX (Table S2). The content of F and Cl are displayed at the same time in Table S2 and Table S3. It should be noted that carbon pollution is inevitable in XPS and EDX measurement, so the content of C in Table S2 and Table S3 is for reference only. Based on the structural and elemental analysis results, we successfully prepared $Nb_2CF_x$ and $Nb_2CCl_x$ MXenes.

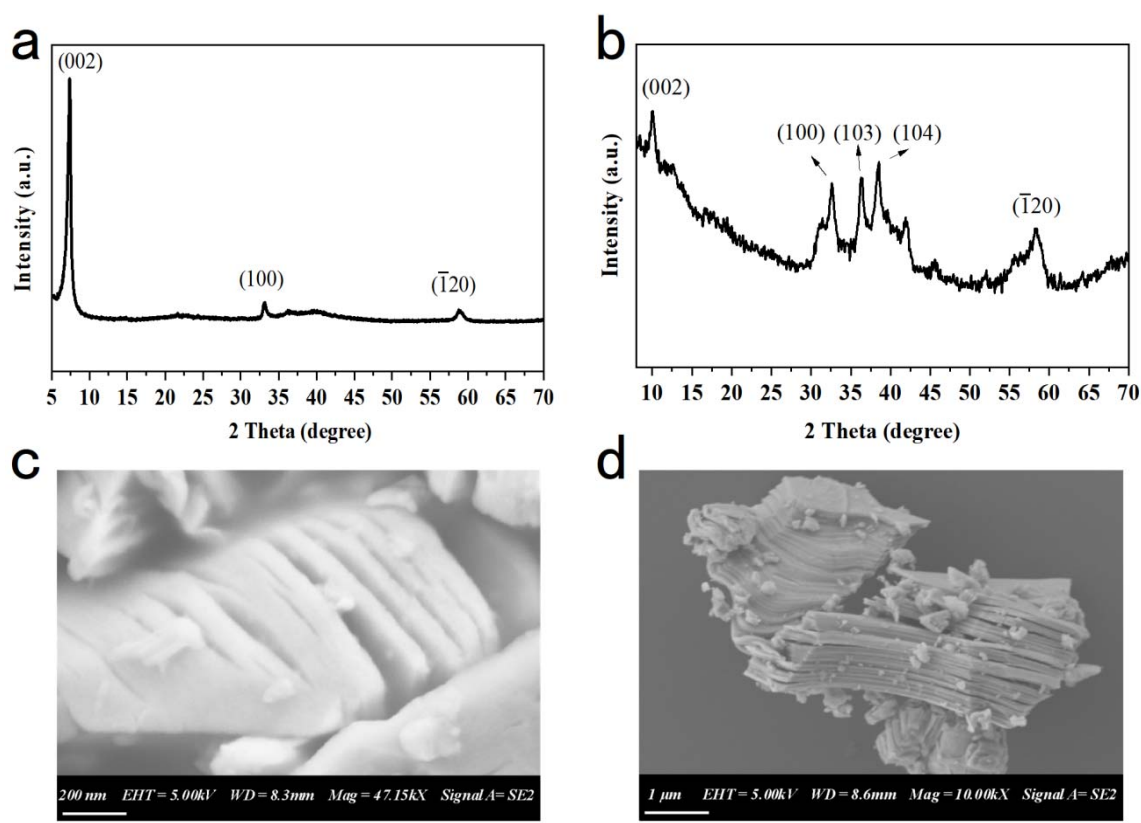

**Fig. 1.** (a) The XRD pattern of $Nb_2CF_x$ prepared by the hydrothermal method, (b) The XRD pattern of the prepared $Nb_2CCl_x$, The SEM images of (c) $Nb_2CF_x$ and (d) $Nb_2CCl_x$.

3.2 Magnetic and electrical measurements

The Meissner effect and zero resistivity are the two most basic and prerequisite features of superconductivity. To study the effect of two different functional groups on the superconductivity of $Nb_2C$-MXene, we measured the dependent curves of magnetization with temperature on $Nb_2CF_x$ and $Nb_2CCl_x$. The powder sample was put in a plastic capsule with a linear diamagnetic background to measure its magnetization. Figure 2a shows the magnetization versus temperature of $Nb_2CF_x$ synthesized from the hydrothermal method measured at 20 Oe, which does not exhibit superconducting behavior. The temperature dependence of the magnetization of $Nb_2CCl_x$ with zero-field cooling (ZFC) and field cooling (FC) at 20 Oe is illustrated in Fig. 2b. It is apparent that the magnetization of $Nb_2CCl_x$ has a sudden diamagnetic drop at 5.2 K, and the ZFC and FC curves are bifurcated at this temperature. According to the Meissner effect, it can be judged that the $Nb_2CCl_x$ is superconducting, and the $T_c$ is ~ 5.2 K, comparable to the value of 6 K reported by Kamysbayev et al. [21]. Given the crystallographic density is 5.3 g/cm$^3$, from the magnetic susceptibility at 2 K of -0.00342 emu/(g·Oe), the shielding volume fraction is estimated as 22.8%.

Our experimental results show that $Nb_2CF_x$ is not superconducting, which is converse from the recent reports [22, 23]. It should be pointed out that the diamagnetic shielding volume fraction estimated by their magnetic measurements was only about 0.12%, and the zero resistivity ($\rho$) data of the superconductor was absent. To further study the physical properties of $Nb_2CF_x$, it is necessary to measure the resistivity curves with temperature. Figure 2c illustrates the temperature-dependent resistivity of $Nb_2CF_x$ prepared by the hydrothermal method. When the temperature drops below 100 K, the resistivity increasing abruptly, and no superconductivity was detected even the temperature down to 2 K. We also measured the magnetization and resistivity of $Nb_2CF_x$ etched by HF, and the result was similar to that taken from the hydrothermal method, there was no sign of superconductivity (Fig. S3).

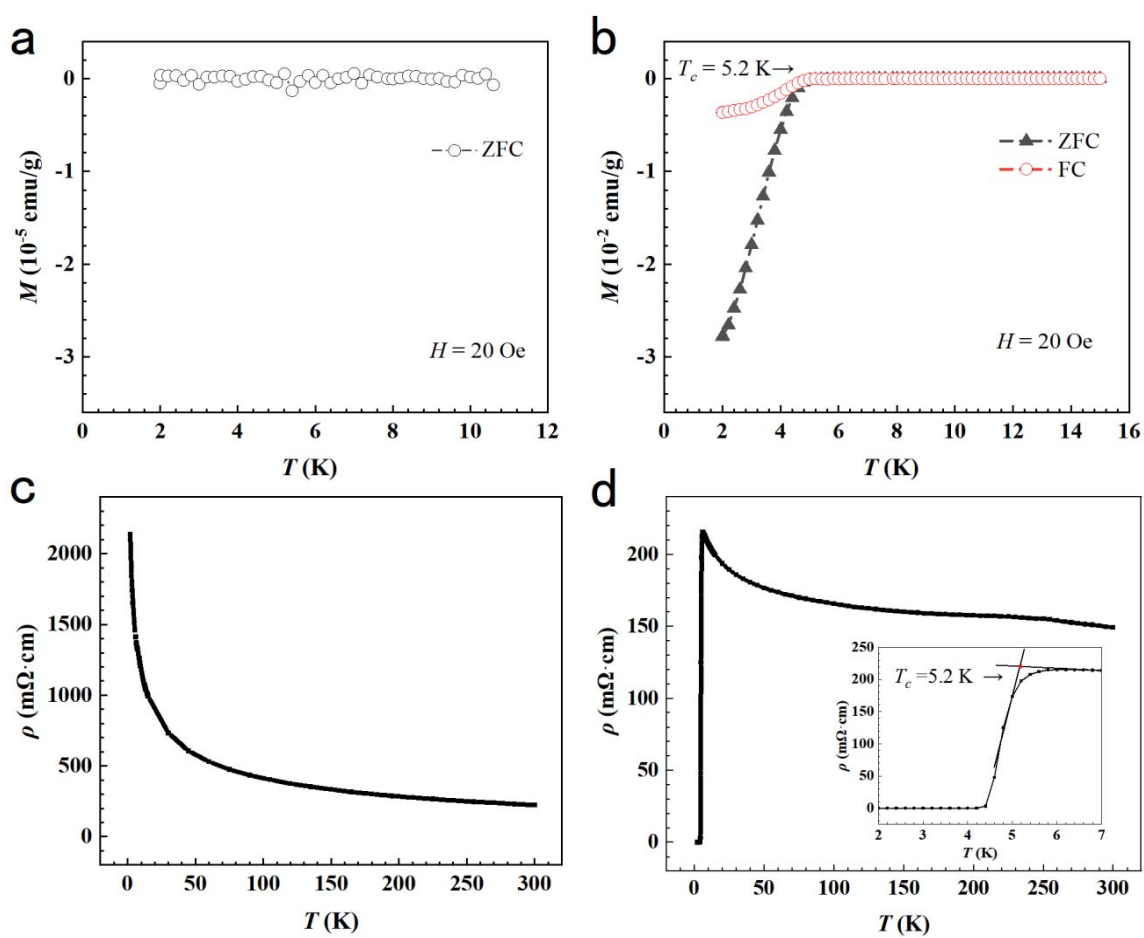

**Fig. 2.** (a) Temperature dependence of magnetization of $Nb_2CF_x$ (Hydrothermally etched) under 20 Oe field, (b) FC and ZFC curves under 20 Oe field of $Nb_2CCl_x$, (c) Resistivity versus temperature curve of $Nb_2CF_x$ (Hydrothermally etched), (d) Temperature dependence of resistivity of $Nb_2CCl_x$. The inset illustrates the data of the low-temperature section and how to define the $T_c$.

As shown in fig. 2d, the resistivity of $Nb_2CCl_x$ increases by about 42% in the temperature range of 300 K to 8 K, indicating that $Nb_2CCl_x$ behaves more like non-metallic. When the temperature is below 6 K, the resistivity drops suddenly and sharply and then stabilizes to zero after 4 K. This provides very firm evidence for the existence of superconductivity of $Nb_2CCl_x$. The inset of Fig. 2d manifests an enlarged view of the $\rho$ - $T$ curve. We define $T_c$ as taking the intersection of the tangent of the two lines of the normal state and the dropping state. It can be seen from the illustration that $T_c$ is ~ 5.2 K, which is similar to data obtained from the magnetic measurement (Fig. 2b).

The AC magnetic susceptibility can provide further information for superconductivity. The real component $\chi_{ac}'$ of the ac susceptibility reveals the diamagnetic shielding, and the imaginary part $\chi_{ac}''$ is a measure of the absorption of energy as the sample transitions between the normal to the superconducting states. As shown in fig. 3, both of them change abruptly at around 5.2 K. The transition temperature is the same as the temperature measured by the ZFC and FC curves. Combining the Meissner effect, the measurement of AC susceptibility and resistivity, we are convinced that $Nb_2CCl_x$ is superconducting with $T_c$ of 5.2 K.

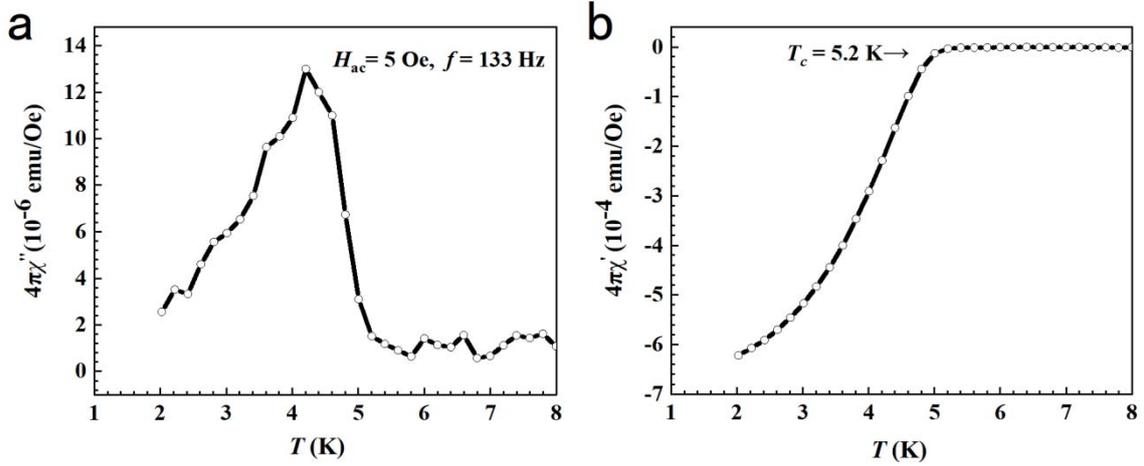

**Fig. 3.** (a) The imaginary magnetic susceptibility $\chi_{ac}''$ and (b) The real magnetic susceptibility $\chi_{ac}'$ with temperature of $Nb_2CCl_x$, indicating a superconducting $T_c$ of ~ 5.2 K. The harmonic magnetic field and frequency are 5 Oe and 133 Hz, respectively.

3.3 Superconducting state parameters of $Nb_2CCl_x$

Generally, we can determine the type of superconductor by the Ginzburg-Landau parameter $\kappa$, which is calculated by the fitting upper critical field ($H_{c2}$) and lower critical field ($H_{c1}$) according to the Ginzburg-Landau theory. To figure out the $H_{c2}$ and $H_{c1}$ of $Nb_2CCl_x$, we measured superconductivity under the various magnetic field at different temperatures. Figure 4 shows the curve of the resistivity versus temperature under different magnetic fields up to 2.8 T. As the magnetic field increases, the $\rho$-$T$ curve shifts to a lower temperature, indicating that superconductivity is suppressed. At a temperature of 2 K, the superconductivity is still not destroyed when the magnetic field is increased to 2.8 T. The $\rho$-$T$ curve under different magnetic fields can determine the superconducting parameter $H_{c2}$. The $T_c$ used to determine $H_{c2}$ is the same as the previous $T_c$ criterion for resistivity measurement. The inset of Fig. 4 indicates the measured value of $H_{c2}$ and the fitting curve (solid line) according to the Ginzburg-Landau equation [32]:

$$H_{c2}(T) = H_{c2}(0)\left[1-\left(T/T_c\right)^2\right] \quad (1)$$

The obtained value of $H_{c2}(0)$ is 3.57 T.

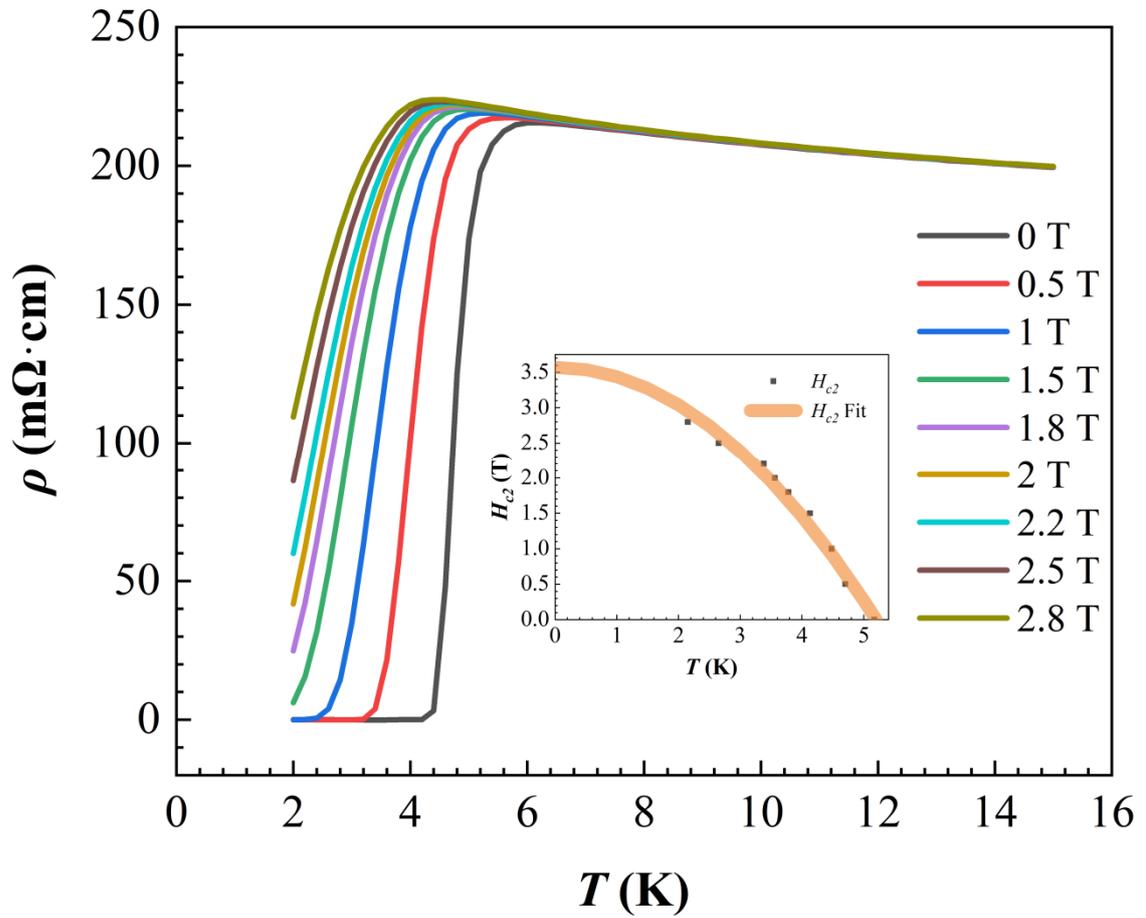

**Fig. 4.** Temperature dependence of resistivity at various magnetic fields from 0.0 to 2.8 T of $Nb_2CCl_x$. The inset illustration shows the experimental value of $H_{c2}$ determined by the $\rho$ - $T$ curve and the fitted curve with the Ginzburg-Landau theory.

Figure 5 shows the magnetic fields dependence of magnetization in the temperature range of 2-4 K. Because there are some paramagnetic impurities in the tested samples, we have made background corrections to the data. The butterfly curves of hysteresis loops depict the typical characteristic of type-II superconductors. In the beginning, the magnetization decreases linearly until the magnetic field exceeds the lower critical field ($H_{c1}$). The inset of Fig. 5 presents the change of $H_{c1}$ and $H_{c2}$ with temperature and the fitting curve (solid line) according to the Ginzburg-Landau equation:

$$H_{c2}(T) = H_{c2}(0)\left[1-\left(T/T_c\right)^2\right] \tag{2}$$

$$H_{c1}(T) = H_{c1}(0)\left[1-\left(T/T_c\right)^2\right] \tag{3}$$

The obtained values of $H_{c1}(0)$ and $H_{c2}(0)$ are 0.283 T and 3.3 T.

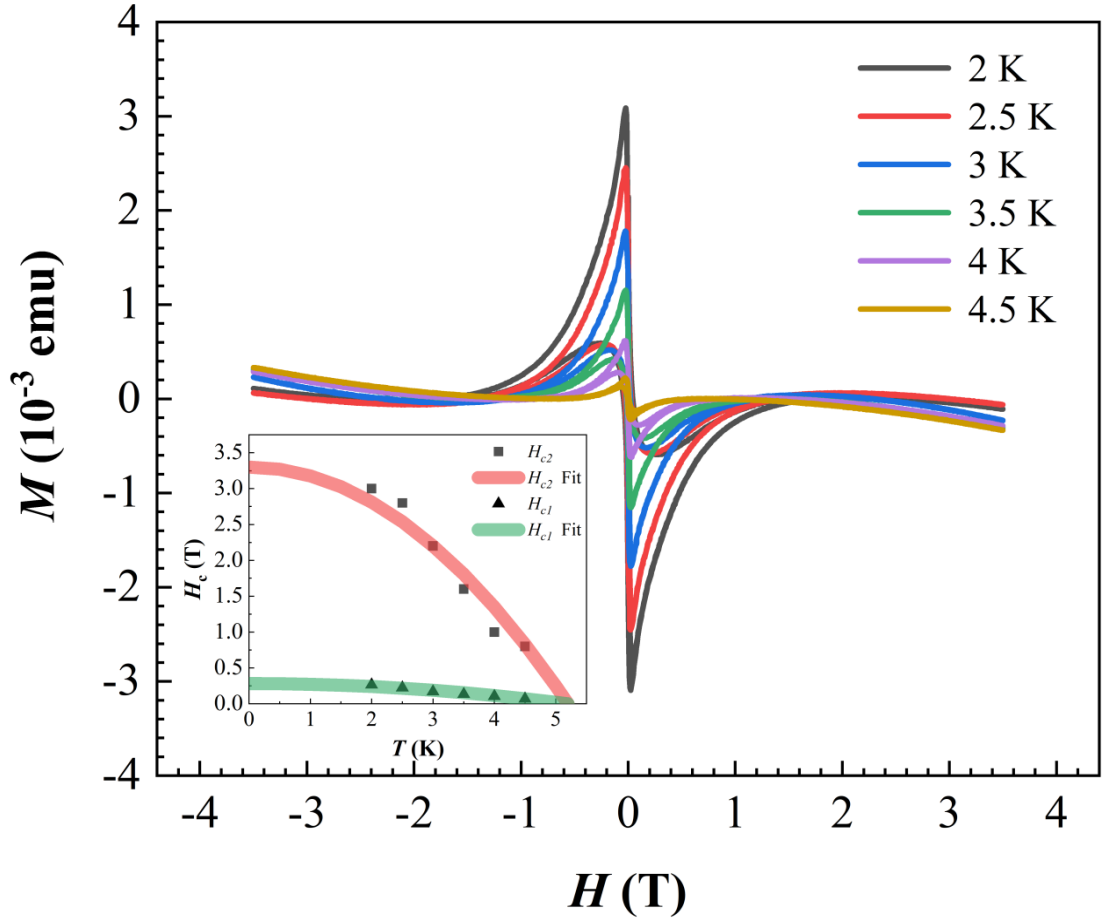

**Fig. 5.** Magnetization (*M*) dependence on the applied magnetic field (*H*) hysteresis loops measured at different temperatures of $Nb_2CCl_x$. The inset illustration shows the experimental value of $H_{c1}$ and $H_{c2}$ determined by the hysteresis curve and the fitted curve (solid line) with the Ginzburg-Landau theory.

The coherence length $\xi_{GL}$ and penetration depth $\lambda_{GL}$ can be determined by the equation for the critical fields:

$$H_{c2}(0) = \frac{\Phi_0}{2\pi \xi_{GL}^2} \tag{4}$$

$$H_{c1}(0) = \frac{\Phi_0}{4\pi \lambda_{GL}^2} \ln \frac{\lambda_{GL}}{\xi_{GL}} \tag{5}$$

Where $\Phi_0 = \frac{h}{2e}$ is the flux quantum, $H_{c2}(0)$ is the Ginzburg-Landau fitting data from the $\rho$-$T$ curve. In this way, we obtained the coherence length $\xi_{GL}$ = 96.08 Å and penetration depth $\lambda_{GL}$ = 231.36 Å.

The superconducting thermodynamic critical field $H_c(0)$ can be determined by the geometric mean of $H_{c1}(0)$ and $H_{c2}(0)$ as $H_c(0) = \sqrt{H_{c1}(0) \times H_{c2}(0)}$ [32], the calculated value is to be 0.97 T. According to Ginzburg-Landau theory, $H_{c2}$ and $H_c$ have the following relationship: $H_{c2} = \sqrt{2}\kappa_{GL}H_c$. Thus, the Ginzburg-Landau parameter $\kappa_{GL}$ is estimated to be 2.41. This parameter also confirms that Nb$_2$CCl$_x$ is a type-II superconductor. The superconducting parameters of Nb$_2$CCl$_x$ are summarized in Table 1.

Table 1. Superconducting parameters of Nb$_2$CCl$_x$

| Parameter | Unit | Value |
| --- | --- | --- |
| $T_c$ | K | 5.2 |
| $H_{c1}$ | T | 0.28 |
| $H_{c2}$ | T | 3.57 |
| $H_c$ | T | 0.99 |
| $\xi_{GL}$ | Å | 96.08 |
| $\lambda_{GL}$ | Å | 231.36 |
| $\kappa_{GL}$ | --- | 2.41 |

3.4 The effect of structure and localization on superconductivity

Our experimental results clearly show that Nb$_2$CCl$_x$ is superconducting while Nb$_2$CF$_x$ is not. In addition to the different functional groups, it can be found from

Fig.1ab and Fig. S1 that the layer distances of the three samples are different. It is well established that annealing can reduce layer spacing of MXenes [33, 34]. To verify whether the layer spacing has an effect on the superconductivity, we annealed $Nb_2CF_x$ under a vacuum at 250 °C. It was shown that the interlayer spacing of $Nb_2CF_x$ was reduced from 11.98 Å to 8.82 Å after annealing, which was similar to the value of $Nb_2CCl_x$. The XRD patterns before and after annealing are illustrated in Fig. S4a, only the (002) peak is shifted and no other impurities are introduced. However, magnetic measurement shows the annealed $Nb_2CF_x$ sample is still not superconducting (Fig. S4b). Therefore, we confirm that the functional group is the main reason that affects the superconducting performance, not the layer spacing. Table S1 gives the different structural parameters of $Nb_2CCl_x$ and $Nb_2CF_x$ and annealed $Nb_2CF_x$.

From 10 K to 150 K, we found that the $\rho$-$T$ curves of $Nb_2CF_x$ and $Nb_2CCl_x$ are very similar. This non-metallic behavior can be illuminated as the two-dimensional weak localization (WL) of charge carriers caused by the disorder. We fitted the resistivity data based on the function:

$$\rho = \frac{1}{\sigma_0 + a T^{\frac{1}{2}}} + b T^2 \qquad (6)$$

where $\sigma_0 = 1/\rho_0$ is the residual conductivity, $aT^{1/2}$ is related to quantum correction from the electron-electron interaction, and the second term describes the high temperature part [35-39]. As shown in Fig. 6, this model describes the two experimental data very well ($R^2 = 0.99$) and give the fit parameters for $Nb_2CF_x$: $\sigma_0 =$ 1.364×10$^{-4}$ mΩ$^{-1}$·cm$^{-1}$, $a$ = 2.247×10$^{-4}$ mΩ$^{-1}$·cm$^{-1}$·K$^{-1/2}$, and $b$ = -6.065×10$^{-4}$

mΩ·cm·K$^{-2}$; and for Nb$_2$CCl$_x$: $\sigma_0$ = 4.17×10$^{-3}$ mΩ$^{-1}$·cm$^{-1}$, $a$ = 2.15×10$^{-4}$ mΩ$^{-1}$·cm$^{-1}$·K$^{-1/2}$, and $b$ = 6.55×10$^{-4}$ mΩ·cm·K$^{-2}$. It should be noted that magnetoresistance measurements on Nb$_2$C-MXene etched by hydrofluoric acid also manifested similar weak localization behavior [40].

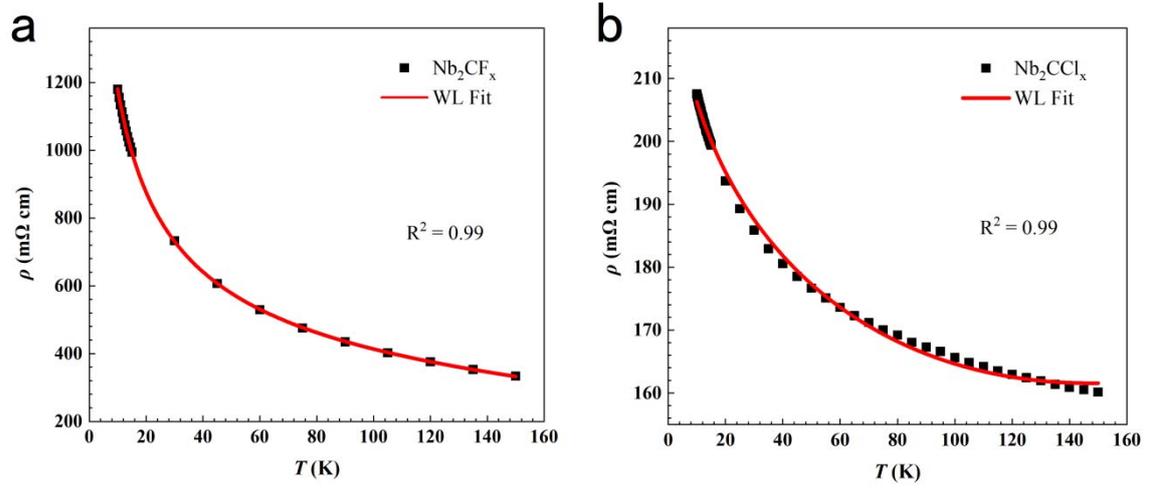

**Fig. 6.** The WL fitting curve of resistivity of (a) Nb$_2$CF$_x$, and (b) Nb$_2$CCl$_x$.

More importantly, parameter $a$ is related to the disorder, which is enhanced with the increase of the disorder increases [41]. Meanwhile, the weak localization effect weakens electron-phonon interactions, that is to say, superconductivity is suppressed [42]. The parameter $a$ follows the relation:

$$a = \lambda_F^4 \left( \frac{ne^2}{h} \right)^{3/2} \frac{(2m^*k_B)^{1/2}}{h} \sigma_0^{-1/2} \qquad (7)$$

$$\lambda_F \propto N(E_F) \qquad (8)$$

The density of states on the Fermi surface ($N(E_F)$) directly affects the formation of Cooper pairs, which is the premise of superconductivity. Taking the obtained WL fitting parameters into formula (7), we draw a conclusion that $\lambda_F$ of Nb$_2$CF$_x$ is smaller than $Nb_2CCl_x$. Considering $N(E_F)$ is directly proportional to $\lambda_F$ (formula (8)), $N(E_F)$ of $Nb_2CCl_x$ is more than that of $Nb_2CF_x$. Our following theoretical calculations also support the conclusion. Thus, the functional group of $Nb_2C$-MXene will result in different extents of carriers localization and significantly change $N(E_F)$, which may be the main reason why $Nb_2CCl_x$ is superconducting, whereas $Nb_2CF_x$ is not.

3.5 Theoretical calculations

To get an insight into the superconductivity of $Nb_2C$-MXene, we performed DFT calculations using the QUANTUM-ESPRESSO package, which was widely used to compute the surface properties [43-46]. In order to reduce the computational cost, we adopted a structure nominally composed of $Nb_2CCl_2$ ($Nb_2CF_2$) [21], which was indexed to the space group $P6_3/mmc$ with optimized lattice constants $a$ = 3.33 (3.20) Å and $c$ = 19.12 (17.00) Å, respectively (Fig. 7a). Fig. 7b show the first Brillouin zone of $Nb_2CCl_2$ and the Highly symmetric paths Γ→K→M→A. This structure is dynamically stable, as no imaginary phonon frequency appears in the Brillouin zone (Fig. 7c). We calculate the $T_c$ by using the McMillan-Allen-Dynes parameterized Eliashberg equation [47],

$$T_c = \frac{\omega_{\log}}{1.2} \exp\left(-\frac{1.04(1+\lambda)}{\lambda - \mu^*(1+0.62\lambda)}\right) \quad (9)$$

where $\omega_{\log}$ is the logarithmic averaged phonon frequency and $\lambda$ is the total electron-phonon coupling constant.

Figure S5 illustrates the range of $T_c$ variation with Coulomb repulsion pseudopotential $\mu^*$. The calculated value of $T_c$ decreases with increasing $\mu^*$. An estimated value of $T_c$ is 5.3 K with $\mu^* = 0.1$, which is a reasonable value used widely in the literature [48-54]. The results show that $Nb_2CCl_2$ is a BCS superconductor. Phonon density of states (PHDOS) and Eliashberg function $\alpha^2F(\omega)$ for $Nb_2CCl_2$ are shown in Figs. 7d and 7e. Their peaks almost coincide. The densities of state of low frequency modes accounts for most of PHDOS. The low frequency vibration modes below 200 cm$^{-1}$ induce the main electron-phonon coupling. The electron-phonon coupling constant $\lambda$ is 0.63. Phonon softening has a great effect on superconductivity. The noticeable low frequency soft phonons at $\frac{1}{2}$ ΓK with significant big linewidth (Fig. 7c) led to the peak of Eliashberg spectral function.

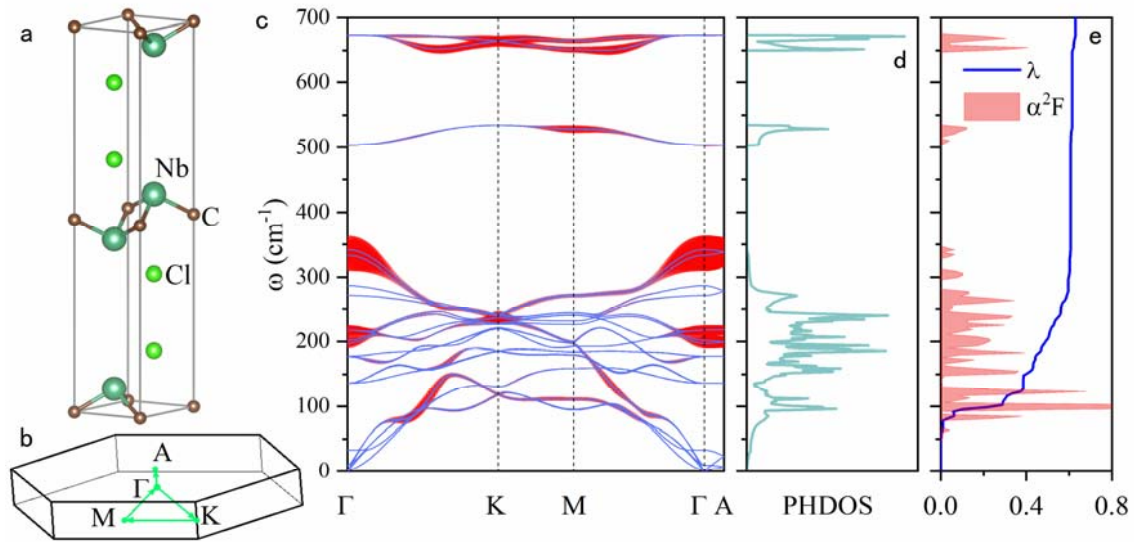

**Fig. 7.** (a) Crystal structure of $Nb_2CCl_2$ in unit cell with space group $P6_3/mmc$, (b) The first Brillouin zones, (c) Phonon dispersion with phonon linewidth $\gamma_{qv}$ in red bubble, (d) Phonon density of states, (e) Eliashberg function $\alpha^2F(\omega)$ and $\lambda(\omega)$.

Rather, our calculations show that the $Nb_2CF_2$ with space group $P6_3/mmc$ is dynamically unstable. As shown in Fig. S6, the phonon mode with significant imaginary frequency at high symmetric point M indicates that the $Nb_2CF_2$ prefers to form (2×2×1) CDW phase than the normal phase. The calculated electronic structures show that the $N(E_F)$ of $Nb_2CCl_2$ of 3.2 states/eV/cell is larger than that of $Nb_2CF_2$ of 2.6 states/eV/cell (Fig. S7 and S8), which is consistent with our previous transport experimental analysis and may dominate whether or not superconductivity occurs. The present studies thus should give valuable insight into the precise ground state structure of $Nb_2C$-MXene and the surface-groups- dependent superconductivity.

## 5. Conclusion

In summary, the role of surface functional groups to superconductivity in $Nb_2C$-MXene has been investigated by experiments and DFT calculations. We have corroborated that $Nb_2C$-MXene with Cl functional group is superconducting and belongs to the type-II superconductor, while $Nb_2CF_x$ with F functional group is not superconducting. The different functional groups of $Nb_2C$-MXene affect the carrier localization and significantly change the density of states on the Fermi surface, which may lead to superconductivity emergence or not. What needs illustration is that these two $Nb_2C$-MXenes were synthesized by different methods. It would be highly interesting to fabricate $Nb_2CF_x$ by the high-temperature molten salts method and examine its superconductivity. Our results clarify the inconsistencies in the current researches and provide an insight into the superconductivity of MXenes materials.

**Declaration of Competing Interest**

We declare that we do not have any known competing financial interests or personal relationships that could have appeared to influence the work reported in this paper.

**Acknowledgements**

This work was supported by the Key Technologies R&D Program of Guangzhou City (No. 201704020182 and 201803030008), Guangzhou Basic and Applied Basic Research Foundation (No. 202102080166), and Open Foundation of Guangxi Key Laboratory of Processing for Non-ferrous Metals and Featured Materials, Guangxi University (No. 2021GXYSOF11).

**Supplementary materials**

Supplementary material associated with this article can be found, in the online version, at doi:10.1016/j.surfin.XXXX.XXXXXX.